\documentclass[12pt]{article}
\usepackage{epsfig}

\def\beq{\begin{equation}}
\def\eeq#1{\label{#1}\end{equation}}
\def\eeqn{\end{equation}}

\def\beqa{\begin{eqnarray}}
\def\eeqa#1{\label{#1}\end{eqnarray}}
\def\eeqan{\end{eqnarray}}

\let\bar=\overbar

\def\Dslash{\not{\hbox{\kern-4pt $D$}}}
\def\dslash{\not{\hbox{\kern-2pt $\del$}}}

\def\msb{{\bar{\ssstyle M \kern -1pt S}}}

\usepackage{fancyhdr,graphicx}
\def\Title#1{\begin{center} {\Large {\bf #1} } \end{center}}

\begin{document}

\Title{From field theory to superfluid hydrodynamics of dense quark matter}

\bigskip\bigskip


\begin{raggedright}
Mark G. Alford, S. Kumar Mallavarapu \\
{\it Department of Physics, Washington University St Louis, MO, 63130, USA\\  
}

\bigskip
Andreas Schmitt, Stephan Stetina\footnote{Speaker} \\
{\it Institute for Theoretical Physics, Vienna University of Technology, 1040 Vienna, Austria\\
}
\bigskip\bigskip
\end{raggedright}

\section{Introduction}

The study of quantum chromodynamics (QCD) at low temperatures and
high densities is relevant for fundamental as well as applied astrophysical
questions. On the fundamental side, the phase diagram of QCD at high
and intermediate densities might be very complicated and its study
is highly non-trivial. While heavy-ion collisions and lattice calculations
are powerful tools to probe the phase structure in a regime of high
temperatures and low densities, their applicability at higher densities
is very limited. However, future accelerator facilities such as NICA might
provide some insight (see for example Ref.\ \cite{Dubna}).

Reliable information can be obtained in a region of asymptotically
high densities, where QCD behaves like a free field theory and perturbative
calculations are possible. The ground state
in this region of the phase diagram is a color superconductor where
quarks of all color and flavor form cooper pairs which was termed
color-flavor locking (CFL) \cite{Alford1}. Among many interesting features
of CFL that could be discussed at this point, two are particularly
important for the following analysis: CFL spontaneously breaks chiral symmetry which
leads to an octet of Goldstone bosons and furthermore, CFL spontaneously
breaks baryon conservation $U(1)_{B}$. Strictly speaking, perturbative
calculations of for example the magnitude of the superconducting gap,
are only valid at (rather exotic) values of the chemical potential
of about $\mu\simeq10^{8}\, {\rm MeV}$ and it is thus questionable whether CFL will
persist all the way down to the phase boundary of nuclear
matter. Going down in density, the increase in the mass of the strange
quark will act as an external stress on the highly symmetric pairing
pattern and systematic studies show \cite{Schaefer}, that CFL will most likely
react by developing a kaon condensate. The corresponding ground state
CFL-$K^{0}$ is then subject to two symmetry breaking patterns: The breaking of
chiral symmetry and baryon conservation in CFL and the breaking of strangeness conservation 
due to the kaon condensate. 
\begin{equation}
SU(3)_{L}\otimes SU(3)_{R}\otimes SU(3)_{C}\otimes U(1)_{B}\rightarrow SU(3)_{L+R+C}\otimes\mathcal{Z}_{2}\,,\,\, SU(3)_{L+R+C}\supseteq U(1)_{S}\rightarrow1\,.\end{equation}
One should note at this point that weak interactions explicitly violate
strangeness conservation and thus $U(1)_{S}$ is not an exact symmetry
to begin with (we shall come back to this point in the outlook). Apart
from CFL-$K^{0}$, there are many other possible candidate
ground states of QCD at intermediate densities \cite{review}, but the absence of controlled experiments in laboratories or reliable
models to perform calculations make it very difficult to decide which phases are realized in nature. 

The study of compact stars as the only ``laboratory`` where such intermediate densities are
realized could prove to be very useful. The maximum value for the chemical potential inside a compact star
is estimated to be high enough for deconfined quark matter to be conceivable
($\mu_{star}\leq500\, {\rm MeV}$) . Usually, most insights into the interior
of compact stars can be obtained from the characteristics of cooling
and neutrino emissivity as well as from mass-radius relations. Other
observable phenomena such as pulsar glitches or r-mode instabilities require a hydrodynamic description. In order to provide a
framework for such calculations, it is important to
understand how an effective hydrodynamic description emerges from
the underlying microscopic physics discussed above. Moreover, different (equivalent) formulations of superfluid hydrodynamics exist in the literature and 
should be connected to each other. These were the central aims of Ref.\ \cite{Alford3} whose main results shall be reviewed here.

\section{Superfluid hydrodynamics from field theory}

The spontaneous breaking of a $U(1)$ symmetry, be it the breaking
of $U(1)_{B}$ or $U(1)_{S}$ discussed above, is known to be the
fundamental mechanism for superfluidity. We will focus here on kaon
condensation (and thus on the breaking of $U(1)_{S}$), but similar
considerations hold for  essentially
 any system with a spontaneously broken symmetry.
 The existence of a condensate together with the absence
of elementary excitations which could dissipate energy allow for frictionless
transport of the associated $U(1)_{S}$ charge at low temperatures.
This might be puzzling at first since it is also well known that
the spontaneous breaking of a global symmetry leads to existence of
massless excitations (Goldstone bosons). At least at zero temperature,
this apparent contradiction can easily be resolved: since the low-energy
dispersion relation of the Goldstone mode is linear in momentum, it
can only be excited beyond a certain critical velocity which is given
by the slope of the linear part of the dispersion. At nonzero temperature,
the situation is more complicated as thermal excitations of the Goldstone
mode ({}``phonons'') are present for any superfluid velocity. As
a consequence, in the translation into hydrodynamic equations, we
will have to consider both ingredients, condensate and phonons, and they also appear coupled to each other. 

The hydrodynamic framework capable of describing superfluids, which
we will derive now from microscopic physics, was first set up by Landau
\cite{Landau} mainly for the (non-relativistic) purpose
of describing superfluid helium. The basic idea is to formally divide
the fluid into a superfluid and normal fluid part. 
The total mass density (later in the relativistic context to be replaced
by charge density) is then given by $\rho=\rho_{n}+\rho_{s}$ where
at zero temperature only the superfluid density is present and above
the critical temperature only the normal fluid density.  From this point of view, the hydrodynamic description of CFL
with kaon condensation is quite complex: due to the breaking of $U(1)_{B}$
and $U(1)_{S}$ two superfluid components are present such that in
total we expect to be dealing with four different fluid components.
In order to derive the two component hydrodynamics for the kaon part,
we start by analyzing a complex scalar $\varphi^{4}$ model, which
can be interpreted as an effective theory for kaons \cite{Alford4}. 

\subsection{Zero temperature}

Based on Ref.\ \cite{Son1}, we start with our derivation at zero temperature, where only the superfluid component is present. As mentioned at
the end of the last section, our starting point will be a $\varphi^{4}$
theory where the complex doublet field can be interpreted as $(\varphi,\varphi^{*})=(K_{0},\bar{K}_{0})$: 
\begin{equation}
\mathcal{L}=\partial_{\mu}\varphi\partial^{\mu}\varphi^{*}-m^{2}\left|\varphi\right|^{2}-\lambda\left|\varphi\right|^{4}\,\,.\end{equation}
After introducing the condensate,
\begin{equation}
{\displaystyle \varphi\rightarrow\left\langle \varphi\right\rangle + {\rm fluctuations} \, , \qquad  
\left\langle \varphi\right\rangle =\frac{\rho(x)}{\sqrt{2}}e^{i\psi(x)}}\,\,,\end{equation}
we simplify calculations by additionally assuming a constant modulus
$\rho$ of the condensate. The equations of motion then imply \cite{Alford3}
that also $\partial^{\mu}\psi$ is constant and the tree-level potential turns into:
\begin{equation}
U\equiv-\mathcal{L}=-\frac{(\sigma^{2}-m^{2})^{2}}{4\lambda}\,\,,\,\,\,\,\,\,\,\,\,\,\,\,\,\sigma\equiv\sqrt{\partial_{\mu}\psi\partial^{\mu}\psi}\,\,.\end{equation} 
Field-theoretic definitions for the Noether current and stress energy
tensor can be matched to the corresponding hydrodynamic expressions: 
\begin{eqnarray}
j^{\mu} & = & n_{s}v_{s}^{\mu}=\frac{\partial\mathcal{L}}{\partial(\partial_{\mu}\psi)}=\partial^{\mu}\psi\frac{\sigma^{2}-m^{2}}{\lambda}\,\,,\\
T^{\mu\nu} & = & (\epsilon_{s}+P_{s})v_{s}^{\mu}v_{s}^{\nu}-g^{\mu\nu}P_{s}=\frac{2}{\sqrt{-g}}\frac{\delta\left(\sqrt{-g}\mathcal{L}\right)}{\delta g_{\mu\nu}}=\partial^{\mu}\psi\partial^{\nu}\psi\frac{\sigma^{2}-m^{2}}{\lambda}-g^{\mu\nu}\mathcal{L}\,\,.\end{eqnarray}
Charge density $n_{s}$ and the flow velocity of the superfluid $v_{s}^{\mu}$ can than easily
be obtained:
\begin{equation}
n_{s}=\sqrt{j_{\mu}j^{\mu}}=\sigma\frac{\sigma^{2}-m^{2}}{\lambda}\,\,,\,\,\,\,\,\,\,\,\,\,\,\, v_{s}^{\mu}=\frac{\partial^{\mu}\psi}{\sigma}\,\,,\end{equation}
as can be energy and pressure density of the superfluid ($\epsilon_{s}$ and $P_{s}$) by making use of proper projections
on $T^{\mu\nu}$. It is important to notice that our simplification
of $\partial^{\mu}\psi$ being constant translates into a uniform
superflow velocity. Finally we include thermodynamics into our
considerations which allows us to identify the chemical potential in 
the superfluid rest frame, $\mu_{s}=\sigma$. The Lorentz invariant quantity $\sigma$ can also be expressed in terms of the 
chemical potential in the normal-fluid rest frame $\mu=\partial_{0}\psi$: with the above definition of the superfluid velocity, a usual Lorentz
factor arises in this relation,
\begin{equation}
\sigma=\mu\sqrt{1-\vec{v}_{s}^{2}}\,\,.
\end{equation}
As we can see, both chemical potential and flow velocity of the superfluid
evolve solely from rotations of the phase of the condensate: rotations
of the phase around the U(1) circle per unit time create the chemical potential
and the rotations per unit length gives rise to the superflow
velocity. 

\subsection{Finite temperature}

In order to introduce finite temperature $T$, we use the simple one-loop
effective action with the tree-level potential $U$ as introduced
before and the inverse propagator $S^{-1}$:
\begin{equation}
\Gamma=-\frac{V}{T}U-\frac{1}{2}\sum_{k}{\rm Tr} \ln\frac{S^{-1}(k)}{T^{2}},\,\,\,\,\,\,\,\,\,\,\, S^{-1}(k)=\left(\begin{array}{cc}
-k^{2}+2(\sigma^{2}-m^{2}) & 2ik\cdot\partial\psi\\
-2ik\cdot\partial\psi & -k^{2}\end{array}\right) \, .
\end{equation} 
We work at small coupling $\lambda\ll1$ and restrict
ourselves to small temperatures (much below the critical temperature
and $T\ll\mu$). Furthermore, any temperature dependence of the condensate
$\rho$ has been neglected. These approximations allow us to obtain
analytic results (generalizations are discussed in the outlook). Finite-temperature
calculations are carried out in the Matsubara formalism, i.e. $k=(k_{0},\vec{k})$,
$k_{0}=-i\omega_{n}$ with bosonic Matsubara frequencies $\omega_{n}=2\pi nT$.
For small momenta, the dispersion of the Goldstone mode obtained from
the poles of the propagator can be written as:
\begin{equation}
\epsilon_{\vec{k}}=c_{1}(x)|\vec{k}|+\frac{c_{2}(x)}{\mu^{2}}|\vec{k}|^{3}+...\,,\end{equation}
and obviously depend on $x=\cos\vartheta$ where $\vartheta$ is the angle
in between $\vec{k}$ and the homogeneous background superflow $\vec{v}_{s}$.
For explicit results for $c_{1}$ and $c_{2}$ please refer
to Ref.\ \cite{Alford3}. 

At finite temperature, a hydrodynamic interpretation of field-theoretic
results is much more challenging. We now have to introduce an additional
{}``normal'' current related to the elementary excitations which
automatically complicates thermodynamics: since our two homogeneous flows
imply two rest frames where either the normal or the super current
is vanishing, it is impossible to define a frame where the pressure
is isotropic. We rather have to deal with pressure densities normal
and longitudinal to the non vanishing current ($T_{\perp}$,$T_{\parallel}$)
in the respective rest frame, but which pressure is the correct
one to appear in the thermodynamic relation? 

\subsubsection{Generalized thermodynamics and entrainment}

The most rigorous answer to this question is contained in the
canonical formalism introduced by Carter \cite{Carter} which provides
an extension of usual thermodynamics by making use of a generalized
pressure $\Psi$ and a generalized energy density $\Lambda$.
The set of variables in this framework differs from the one used in
the two-fluid formalism of Landau: instead of formally dividing the
Noether current into normal and supercurrent, the canonical framework
is based on either the conserved charge and entropy currents $j^{\mu}$
and $s^{\mu}$ (the latter is conserved only in the non-dissipative
case) or their thermodynamically conjugate momenta $\partial^{\mu}\psi$
and $\theta^{\mu}$. Lorentz covariance requires $\Lambda$ and $\Psi$
to be functionals of invariants only (i.e. $\Lambda=\Lambda[j^{2},s^{2},j_{\mu}s^{\mu}],\,\,\Psi=\Psi[\sigma^{2},\theta^{2},\partial_{\mu}\psi\theta^{\mu}]$)
. Here the definition of $\sigma$ from above has been used. The relation of
$\partial^{\mu}\psi$ to the chemical potential has already been discussed
in the previous section and as we show later, $\theta^{\mu}$ can
be related to the temperature. It might seem puzzling
at first to generalize scalar quantities such as temperature to a
four-vector, but the actually measured temperature is of course obtained
by contraction with the velocity four-vector pointing to the rest
frame of interest. We can now obtain the correct equations of motion
by varying either the generalized energy density with respect to the
conserved currents or the generalized pressure with respect
to the conjugated momenta:
\begin{equation}
d\Lambda=\partial_{\mu}\psi dj^{\mu}+\theta_{\mu}ds^{\mu}\,\,,\,\,\,\,\,\,\,\,\,\,\, d\Psi=j_{\mu}d(\partial^{\mu}\psi)+s_{\mu}d\theta^{\mu}\,\,.\end{equation}
Switching between these two descriptions is identical to a change
from a Hamiltonian to a Lagrangian framework, $\Lambda$ and $\Psi$
are connected by a Legendre transformation. With all these ingredients
at hand, we can now write down a generalized thermodynamic relation: 
\begin{equation}
\Lambda+\Psi=j\cdot\partial\psi+s\cdot\theta\,\,.
\end{equation}
Applying the chain rule to Eq.\ (11) allows us to write:
\begin{equation}
\partial^{\mu}\psi=\frac{\partial\Lambda}{\partial j_{\mu}}={\cal B}j^{\mu}+{\cal A}s^{\mu}\,\,,\,\,\,\,\,\theta^{\mu}=\frac{\partial\Lambda}{\partial s_{\mu}}={\cal A}j^{\mu}+Cs^{\mu}\,\,,\end{equation}
with
\begin{equation}
{\cal A}=\frac{\partial\Lambda}{\partial(j\cdot s)}\,\,,\,\,\,\,\,{\cal B}=2\frac{\partial\Lambda}{\partial j^{2}}\,\,,\,\,\,\,\,\,{\cal C}=2\frac{\partial\Lambda}{\partial s^{2}}\,\,.\end{equation}
The coefficient
${\cal A}$ appears in the relations for $\partial^{\mu}\psi$
as well as $\theta^{\mu}$. It is called ``entrainment coefficient''
which reflects the fact that any change in either of the two conjugate
momenta will automatically affect both currents (and the other way
around) such that the currents are coupled to each other. Finally,
the stress-energy tensor is now given by
\begin{equation}
T^{\mu\nu}=-g^{\mu\nu}\Psi+j^{\mu}\partial^{\nu}\psi+s^{\mu}\theta^{\nu}\,\,.\end{equation}
Plugging in Eqs.\ (13), one can see right away that $T^{\mu\nu}$
becomes symmetric in the Lorentz indices. 

It remains to relate the canonical formalism to the original two fluid
picture of Landau (a relativistic generalization of which
can for example be found in \cite{Khala,Son2}). After
introducing a normal-fluid current, the straightforward extension
of our zero-temperature expressions for Noether current and stress-energy
tensor are
\begin{eqnarray}
{\displaystyle T^{\mu\nu}} & = & {\displaystyle \left(\epsilon_{n}+P_{n}\right)u_{n}^{\mu}u_{n}^{\nu}+\left(\epsilon_{s}+P_{s}\right)\partial^{\mu}\psi\partial^{\nu}\psi/\sigma^{2}+(P_{n}+P_{s})g^{\mu\nu}\,\,,}\\
{\displaystyle j^{\mu}} & = & {\displaystyle n_{n}u_{n}^{\mu}+n_{s}\partial^{\mu}\psi/\sigma\,\,.}\end{eqnarray}
Here, we have introduced the flow velocity of the normal fluid $u_{n}^{\mu}$, as well as the corresponding energy and pressure densities $\epsilon_{n}$ and $P_{n}$.
The superfluid velocity is still given by Eq.\ (7) . The coefficients
${\cal A},\,{\cal B}$ and ${\cal C}$ prove to be very useful in
translating one formalism into the other as all hydrodynamic parameters in either model can be expressed
in terms of them:
\begin{equation}
n_{s}=\frac{\sigma}{{\cal B}}\, , \,\,\,\,\,\,n_{n}=-\frac{{\cal A}}{{\cal B}}s,\,\,\,\,\,\,\epsilon_{s}+P_{s}=\frac{\sigma^{2}}{{\cal B}},\,\,\,\,\,\,\epsilon_{n}+P_{n}=\frac{{\cal B}{\cal C}-{\cal A}^{2}}{{\cal B}}s^{2}\,\,.\end{equation}
Note that in (16) and (17) $j^{\mu}$ and at the same time $\partial^{\mu}\psi$
appear as parameters such that in the strict sense of the canonical
framework, this has to be considered a {}``mixed'' form. 

\subsubsection{Explicit results at finite temperature}

In the terminology of the Landau formalism, it can be shown \cite{Alford3} that all microscopic calculations 
at finite temperature are carried out in the normal fluid rest frame where we can identify generalized pressure $\Psi$ and
temperature: 
\begin{equation}
\Psi=\frac{T}{V}\Gamma\,,\,\,\,\,\,\,\,\,\,\,\,\,\,\theta_{0}=T\,\,\,\,\,\,\,\,\,\,\,\,\, \mbox{in the normal fluid rest frame.}\end{equation}
\begin{figure}[htb]
\begin{center}
\includegraphics[scale=0.4803]{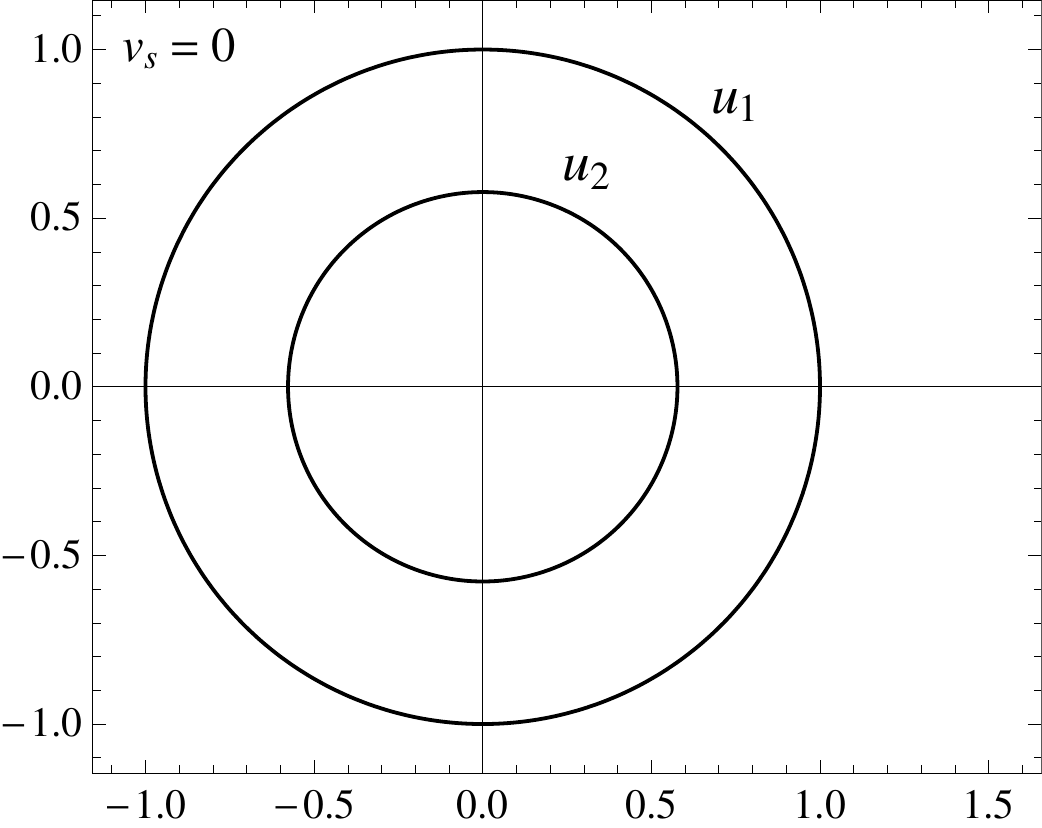}\includegraphics[scale=0.4374]{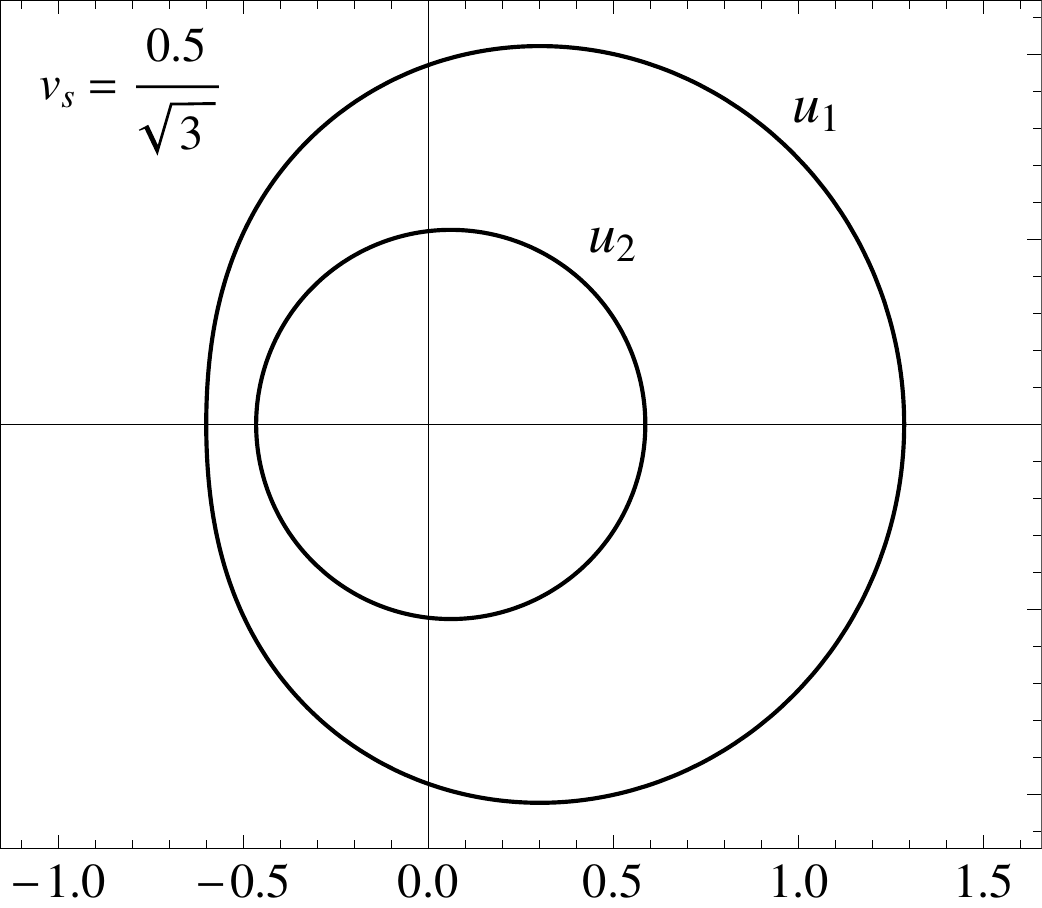}\includegraphics[scale=0.4374]{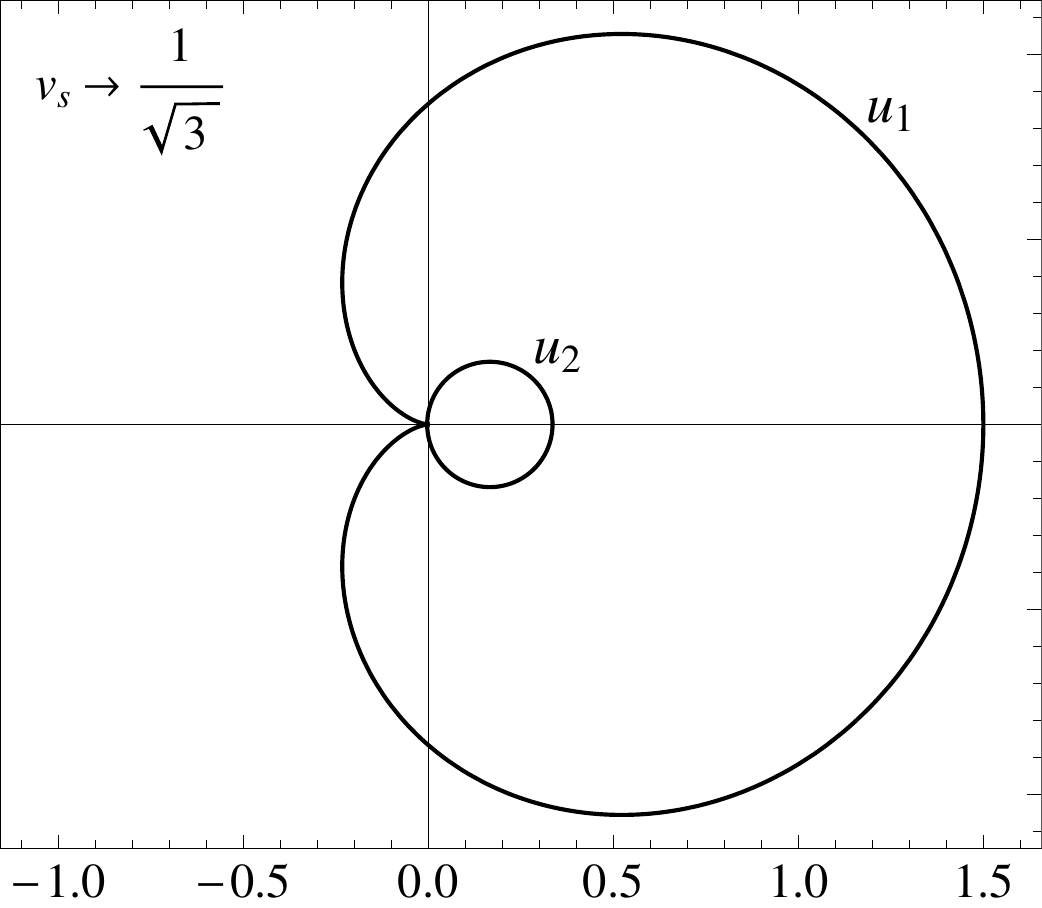}
\caption{Velocities of first and second sound $u_{1}$ and $u_{2}$. The
homogeneous background superflow is aligned along the horizontal axis
and the speeds of sound are calculated for all angles with respect
to the superflow.}
\label{fig:flux}
\end{center}
\end{figure}
Even though the first relation is in principle frame independent, using our microscopic expression (9) automatically restricts us to the normal rest frame. We can now step by step derive
field-theoretic definitions for all the basic quantities of the two-fluid formalism. The coefficients ${\cal A},\,{\cal B}$ and ${\cal C}$ 
are in terms of microscopic variables 
\begin{equation}
{\cal A}=\frac{\partial^{0}\psi}{s^{0}\vec{j}\cdot\vec{\nabla}\psi}\eta\,\,,\,\,\,\,\,\,{\cal B}=-\frac{(\vec{\nabla}\psi)^{2}}{\vec{j}\cdot\vec{\nabla}\psi}\,\,,\,\,\,\,\,\,{\cal C}=-\frac{j^{0}\partial^{0}\psi\,\eta-\vec{j}\cdot\vec{\nabla}\psi s^{0}\theta^{0}}{(s^{0})^{2}\vec{j}\cdot\vec{\nabla}\psi\,\,,}\end{equation}
with the abbreviation $\eta=\vec{v}_{s}^{2}j^{0}\partial^{0}\psi+\vec{j}\cdot\vec{\nabla}\psi$.
With the help of these, we can easily obtain explicit
results  of for example the normal-fluid  density in the low-temperature limit, 
\begin{equation}
n_{n}\simeq\frac{4\pi^{2}}{5\sqrt{3}}\frac{T^{4}}{\mu}\frac{(1-\vec{v}_{s}^{2})^{2}}{(1-3\vec{v}_{s}^{2})^{3}}-\frac{48\pi^{4}}{7\sqrt{3}}\frac{T^{6}}{\mu^{3}}\frac{(1-\vec{v}_{s}^{2})^{2}}{(1-3\vec{v}_{s}^{2})^{6}}\left(15+38\vec{v}_{s}^{2}-9\vec{v}_{s}^{4}\right)\,\,.\end{equation}
As it turns out, the normal fluid density is not just the phonon contribution
to the number density which would yield a $T^{3}$ rather than $T^{4}/\mu$ contribution.

Finally, as an application we present the velocities of first and
second sound in the background of the superflow. The complete calculation
is lengthy and can be found in the appendix of Ref.\ \cite{Alford3}. The final
results for $T=0$ are shown in Fig.\ 1. The sound velocities
depend on the angle between the direction of the sound wave and the
direction of the superflow. The velocity of second sound decreases
significantly as the superflow approaches its critical velocity $v_{crit}=1/\sqrt{3}$.
The impact of small temperature corrections on the speeds of sound is also discussed in Ref.\ \cite{Alford3}. 

\section{Outlook}

For the sake of obtaining analytic results, simplifications were made at several points. This was a necessary and
important first step to gain an understanding of how superfluid hydrodynamics
emerge from the underlying microscopic physics. Restriction to a low-temperature regime can be overcome by making use of the two-particle
irreducible effective action \cite{Alford4,CJT}. All hydrodynamic parameters
can then be calculated numerically up to the critical temperature
(and beyond) and the temperature dependence of the condensate can
be taken into account. Furthermore the explicit breaking of $U(1)_{S}$ due to weak interactions
can be taken into account by adding a small symmetry breaking term
to the Lagrangian.  

\textsl{Acknowledgements:} This work has been supported by the Austrian
science foundation FWF under project no. P23536-N16 and by U.S.~Department of Energy under contract
\#DE-FG02-05ER41375, 
and by the DoE Topical Collaboration 
``Neutrinos and Nucleosynthesis in Hot and Dense Matter'', 
contract \#DE-SC0004955. I thank the organizers for giving me the opportunity to present my work in the frame of this inspiring conference.

\end{document}